\documentclass[ showpacs,showkeys, preprintnumbers,amsmath,amssymb,floatfix,prd]{revtex4}

\usepackage{graphicx}
\usepackage{dcolumn}
\usepackage{bm}

\newcommand{\beq}{\begin{equation}}
\newcommand{\eeq}{\end{equation}}
\newcommand{\bey}{\begin{eqnarray}}
\newcommand{\eey}{\end{eqnarray}}

\begin{document}                             

\title{
Unifying inflation with late-time acceleration by a BIonic system }
\author{Alireza Sepehri}
\email{ alireza.sepehri@uk.ac.ir} \affiliation{Faculty of Physics,
Shahid Bahonar University, P.O. Box 76175, Kerman, Iran}
\author{Farook Rahaman}
\email{rahaman@iucaa.ernet.in} \affiliation{Department of
Mathematics, Jadavpur University, Kolkata 700032, West Bengal,
India}

\author{Mohammad Reza Setare}
\email{ rezakord@ipm.i} \affiliation{Department of Science,
Campus of Bijar, University of Kurdistan, Bijar, \textbf{Iran.}}

\author{Anirudh Pradhan}
\email{ pradhan@iucaa.ernet.in} \affiliation{Department of
Mathematics, Institute of Applied Sciences \& Humanities, GLA University, Mathura-281 406, U.P., India.}

\author{Salvatore Capozziello $^{a,b,c}$}
\email{capozziello@na.infn.it} \affiliation{$^{a }$ Dipartimento
di Fisica, Universit´a di Napoli "Federico II", I-80126 - Napoli,
Italy. \\
$^{ b }$ INFN Sez. di Napoli, Compl. Univ. di Monte S. Angelo,
Edificio G, I-80126 - Napoli, Italy,\\ $^{  c}$ Gran Sasso
Science Institute (INFN), Viale F. Crispi, 7, I-67100, L'Aquila,
Italy}

\author{Iftikar Hossain Sardar}
\email{iftikar.spm@gmail.com} \affiliation{Dept.  of Mathematics,
Jadavpur University, Kolkata 700032, West Bengal, India}

\begin{abstract}
We propose a cosmological model that   unifies inflation, deceleration and
acceleration phases of expansion history by a  BIonic system. At the beginning,
there are $k$ black fundamental strings that transited to the BIon configuration at a given
corresponding point. Here, two coupled universes,  brane and  antibrane, are
created interacting  each other through a wormhole and inflate. With decreasing
temperature, the energy of this wormhole flows into the universe branes and lead to inflation.
After a short time, the wormhole evaporates, the inflation ends and a deceleration epoch starts. By
approaching the  brane and antibrane universes together, a tachyon is born, grows  and causes
the creation of a  new wormhole. At this time, the  brane and antibrane universes result
connected again and the late-time acceleration era of the universe begins. We compare our model
with previous unified phantom models and observational data   obtaining some cosmological
parameters like temperature  in terms of time. We also find that deceleration parameter is
negative during inflation and late-time acceleration epochs, while it is  positive during the deceleration era.
This means that the  model is consistent, in principle,  with  cosmological observations.


\end{abstract}
\maketitle
\section{Introduction}
Recent observations coming from  supernovae surveys, large scale structure and cosmic microwave background radiation  show that the Universe is
presently undergoing a phase of  accelerated phantom expansion
\cite{q1,q2}. Before this era, expansion was decelerated, at least up
to the nucleosynthesis time. This stage of universe history is
well explained by the non-phantom type cosmic fluids. However, another
period of accelerated expansion, named inflation, acts  at very
early epochs  describing  expansion in agreement with   observational
data \cite{q3,q4,q5,q6,q7}. Up to  now, several  models have been
presented to unify the early time inflation with the today observed accelerated phantom phase. For example, some
authors have found that the Universe dynamics begins by an inflationary  phase
and converges towards  a $\Lambda$CDM model if the fluid coupled to dark
energy has a negative energy density at early time \cite{q8}.
Other authors have considered  the recent cosmological deceleration-acceleration
transition redshift in $f(R)$ gravity. They proposed a model where the deceleration parameter
changes sign at a redshift consistent with observations
\cite{q9}. In other scenarios, the future evolution of
quintessence/phantom dominated epoch in modified $f(R)$  gravity has been
considered \cite{odirev,noi}. This type of gravity unifies the early-time inflation
with late-time acceleration and is consistent, in principle,  with observational
data \cite{q10}. Furthermore the universe expansion
history, unifying early-time inflation and late-time acceleration,
can be realized in scalar-tensor gravity minimally or non-minimally coupled to curvature  \cite{q11}.

 However, one of the best models  unifying   the early-time inflation with late-time
acceleration is the phantom cosmology. This model allows to study
the inflationary epoch, the transition to the non-phantom standard
cosmology (radiation/matter dominated eras) and today observed
dark energy epoch. In the unified phantom cosmology,  the same
scalar field plays the role of early time (phantom) inflaton and
late-time Dark Energy. The recent transition from decelerating to
accelerating phase can be also  described  by the same scalar field
\cite{q12}. Despite of this reliable features,  the main question that arises is  on the origin
of the phantom field.  The  answer to this question can come, at a fundamental level, by taking into account a
brane-antibrane system undergoing  three different stages along its evolution. At first stage, $k$ black
fundamental strings  transit  to the so called {\it BIon configuration} at
matching point. The BIon is a configuration in flat space of a
brane universe  and a parallel anti-brane universe connected by a
wormhole \cite{q13,q14}. At transition point, the thermodynamics
of this configuration can be matched to that of $k$ non-extremal
black fundamental strings. At lower temperature, the
wormhole throat becomes smaller, its energy is transferred to the universe
branes and leads to its accelerated expansion. After a short time, this
wormhole evaporates, inflation ends  and non-phantom era begins. This is
the second stage of Universe expansion history. Eventually, two
 brane and antibrane universes become close to each other, the tachyonic
potential between them increases and a new wormhole is formed.
At this stage,  the Universe evolves from the non-phantom phase to the phantom
one and consequently, the late phantom-dominated era starts
and ends up in the big-rip singularity.

We can compare this dynamics
 with the results in  Ref. \cite{q12} and obtain the
wormhole throat features  and temperature in terms of time.

The outline of the paper is  the following.  In Sec. \ref{o1},
we discuss  the inflationary stage in BIon system and show that all
cosmological parameters depend on the wormhole parameters between the
two branes. In Sec. \ref{o2}, we study the second stage where the
wormhole evaporates  and  the pair  brane and
antibrane universes result disconnected. In Sec. \ref{o3}, we consider the third stage where
a new tachyonic wormhole is formed between branes and
accelerates the destruction of the universes towards a big rip. In Sec. \ref{o4}, we
test our model against observational data. The last section is
devoted to summary and conclusions.

\section{ Stage 1: The early time inflation}\label{o1}
In this section, we assume that there is only a fluid of  $k$ black
fundamental strings at  the beginning. In our model, the Universe
is  born at a point corresponding  where the thermodynamics of $k$
non-extremal black fundamental strings is  matched to that of
the BIon configuration. We will construct the inflation in BIon
and discuss that the wormholes between branes have direct effect
on the inflation. We can  also show that all parameters of inflation
depend on the number of branes and on the distance between branes.

Let us  start with the supergravity solution for $k$  coincident
non-extremal black $F$-strings lying along the $z$ direction as discussed in \cite{q14, q15}:
\begin{eqnarray}
&& ds^{2} = H^{-1}(-f dt^{2} + dz^{2})+ f^{-1}dr^{2} + r^{2}d\Omega_{7}^{2},\nonumber\\
&& H = 1 +
\frac{r_{0}^{6}\sinh^{2}\bar{\alpha}}{r^{6}},\:f=1-\frac{r_{0}^{6}}{r^{6}}, \nonumber\\
&& \:
k^{2}=\frac{3^{12}T_{D3}^{4}(\cosh^{2}\bar{\alpha}-1)}{2^{12}\pi^{6}T_{F1}^{2}T^{12}\cosh^{10}\bar{\alpha}}.
\label{Q1}
\end{eqnarray}
In above equation, $T$ is the finite temperature of BIon, $k$ is the
number of black $F$-strings and $T_{D_{3}}$ and $T_{F1}$ are tensions of
brane and fundamental strings respectively. The mass density along
the $z$ direction can be found  from the metric \cite{q15}:

\begin{eqnarray}
&& \frac{dM_{F1}}{dz} = T_{F1}k +
\frac{16(T_{F1}k\pi)^{3/2}T^{3}}{81T_{D3}}+
\frac{40T_{F1}^{2}k^{2}\pi^{3}T^{6}}{729T_{D3}^{2}}.\label{Q2}
\end{eqnarray}
At the corresponding point,  the $k$ black $F$-strings transit to the BIon configuration  where  the string coupling constant ($g_{s}\ll 1$) becomes very small. On the other hand, brane tension depends on the inverse of string coupling (${\displaystyle T_{D3}=\frac{1}{(2\pi)^{3}g_{s}l_{s}^{4}}})$ and tends to larger values at transition point. However, the string tension (${\displaystyle T_{F1}=\frac{1}{2\pi l_{s}^{2}}}$) remains constant and thus $
 {\displaystyle \frac{40T_{F1}^{2}k^{2}\pi^{3}T^{6}}{729T_{D3}^{2}}=\frac{840g_{s}^{2}l_{s}^{4}k^{2}\pi^{7}T^{6}}{729}} $ is  smaller than ${\displaystyle \frac{16(T_{F1}k\pi)^{3/2}T^{3}}{81T_{D3}}=\frac{16g_{s}l_{s}\pi^{3}(2k)^{3/2}T^{3}}{81}}$ and both are  smaller than 1.  Finally, we can write:

\begin{eqnarray}
&& \frac{dM_{F1}}{dz} = T_{F1}k +
A T^{3}+ B T^{6}\nonumber\\
&& A = \frac{16(T_{F1}k\pi)^{3/2}}{81T_{D3}}=\frac{16g_{s}l_{s}\pi^{3}(2k)^{3/2}}{81}\ll 1\nonumber\\&& B=\frac{40T_{F1}^{2}k^{2}\pi^{3}}{729T_{D3}^{2}}=\frac{840g_{s}^{2}l_{s}^{4}k^{2}\pi^{7}}{729}\ll 1\nonumber\\&&\frac{B}{A}\simeq g_{s}l_{s}^{4}\ll 1\label{QQ2}
\end{eqnarray}
Thus, we can ignore higher orders of (${\displaystyle \frac{1}{T_{D3}}}$) in our calculations but the above   approximation is valid.
For finite temperature BIon configurations, the metric takes the form \cite{q14}:
\begin{eqnarray}
&& ds^{2} = -dt^{2} + dr^{2} + r^{2}(d\theta^{2} + \sin^{2}\theta
d\phi^{2}) + \sum_{i=1}^{6}dx_{i}^{2}. \label{Q3}
\end{eqnarray}
 If one chooses  the world volume coordinates of the $D3$-brane as
$\lbrace\sigma^{a}, a=0..3\rbrace$ and defining $\tau =
\sigma^{0},\,\sigma=\sigma^{1}$, then, the coordinates of BIon
assume  the form \cite{q13,q14}:
\begin{eqnarray}
t(\sigma^{a}) =
\tau,\,r(\sigma^{a})=\sigma,\,x_{1}(\sigma^{a})=z(\sigma),\,\theta(\sigma^{a})=\sigma^{2},\,\phi(\sigma^{a})=\sigma^{3}
\label{Q4}
\end{eqnarray}
and the remaining coordinates $x_{i=2,..6}$ are constant. The
embedding function $z(\sigma)$  describes the bending of the
brane. Let $z$ be a transverse coordinate to the branes and $\sigma$
be the radius on the world-volume.  The induced metric on the
brane is:
\begin{eqnarray}
\gamma_{ab}d\sigma^{a}d\sigma^{b} = -d\tau^{2} + (1 +
z'(\sigma)^{2})d\sigma^{2} + \sigma^{2}(d\theta^{2} +
\sin^{2}\theta d\phi^{2}) \label{Q5}
\end{eqnarray}
so that the spatial volume element is ${\displaystyle dV_{3}=\sqrt{1 +
z'(\sigma)^{2}}\sigma^{2}d\Omega_{2}}$. We impose the two boundary
conditions $z(\sigma)\rightarrow 0$ for $\sigma\rightarrow
\infty$ and $z'(\sigma)\rightarrow -\infty$ for $\sigma\rightarrow
\sigma_{0}$, where $\sigma_{0}$ is the minimal two-sphere radius
of the configuration. For this BIon, the mass density along the $z$
direction can be obtained \cite{q14}:
\begin{eqnarray}
&& \frac{dM_{BIon}}{dz} = T_{F1}k + \frac{3\pi T_{F1}^{2}k^{2}
T^{4}}{32T_{D3}^{2}\sigma_{0}^{2}}+
 \frac{7\pi^{2} T_{F1}^{3}k^{3} T^{8}}{512T_{D3}^{4}\sigma_{0}^{4}}.\label{Q6}
\end{eqnarray}
As it can be seen from the above equation, the mass density along the $z$  direction depends on the brane tension ($T_{D3}$).  At transition point, a brane and an antibrane are produced and expand very fast. Consequently, $T_{D3}$ grows and achieve  large values. On the hand, the string tension (${\displaystyle T_{F1}=\frac{1}{2\pi l_{s}^{2}}}$) remains constant and thus ${\displaystyle
 \frac{7\pi^{2} T_{F1}^{3}k^{3} T^{8}}{512T_{D3}^{4}\sigma_{0}^{4}}}$ is  smaller than ${\displaystyle \frac{3\pi T_{F1}^{2}k^{2}
T^{4}}{32T_{D3}^{2}\sigma_{0}^{2}}}$ and both are  smaller than 1. It is

\begin{eqnarray}
&& \frac{dM_{BIon}}{dz} = T_{F1}k +
A' T^{4}+ B' T^{8}\nonumber\\
&& A' =\frac{3\pi T_{F1}^{2}k^{2}
}{32T_{D3}^{2}\sigma_{0}^{2}}=\frac{48\pi^{5} g_{s}^{2}l_{s}^{2}k^{2}
}{32\sigma_{0}^{2}}\ll 1\nonumber\\&& B'=\frac{7\pi^{2} T_{F1}^{3}k^{3} }{512T_{D3}^{4}\sigma_{0}^{4}}=\frac{7\pi^{11}g_{s}^{4}l_{s}^{10} k^{3} }{\sigma_{0}^{4}}\ll 1\nonumber\\&& \frac{B'}{A'}\simeq\frac{1}{T_{D3}^{2}}\simeq g_{s}^{2}l_{s}^{8}\ll 1\label{QQ6}
\end{eqnarray}
For this reason, we can ignore higher order terms in this expression.
Comparing the mass densities for BIon  to the mass density for the
$F$-strings, we see that the thermal BIon configuration behaves like
$k$ $F$-strings at $\sigma = \sigma_{0}$. At this corresponding point,
$\sigma_{0}$ should have the following dependence on the
temperature:
\begin{eqnarray}
&& \sigma_{0} =
\left(\frac{\sqrt{kT_{F1}}}{T_{D3}}\right)^{1/2}\sqrt{T}\left[C_{0} +
C_{1}\frac{\sqrt{kT_{F1}}}{T_{D3}}T^{3}\right].\label{Q7}
\end{eqnarray}
where $T_{F1} = 4k\pi^{2}T_{D3}g_{s}l_{s}^{2}$, $C_{0}$, $C_{1}$,
$F_{0}$, $F_{1}$ and $F_{2}$ are numerical coefficients which can
be determined by requiring that the $T^{3}$ and $T^{6}$ terms in
Eqs. (\ref{Q2}) and (\ref{Q6}) are matched. At this point, the two universes
are  born while  the wormhole is not formed yet. The metric
of these Friedman-Robertson-Walker (FRW)  universes are:
\begin{eqnarray}
&& ds^{2}_{Uni1} = ds^{2}_{Uni2} = -dt^{2} + a(t)^{2}(dx^{2} +
dy^{2} + dz^{2}). \label{Q8}
\end{eqnarray}
The mass density of black $F$-string, BIon and two  universes have to be equal
at the corresponding point:
\begin{eqnarray}
&& \rho_{uni1} + \rho_{uni2} = \frac{dM_{F1}}{dz} =
\frac{dM_{BIon}}{dz} \rightarrow
\nonumber\\
&& -6 H^{2}  = T_{F1}k +
\frac{16(T_{F1}k\pi)^{3/2}T^{3}}{81T_{D3}}+
\frac{40T_{F1}^{2}k^{2}\pi^{3}T^{6}}{729T_{D3}^{2}}, \label{Q9}
\end{eqnarray}
where $H$ is the Hubble parameter. Solving this equation, we obtain:
\begin{eqnarray}
&& a(T) =a(0) e^{-X(T)}, \nonumber\\
&& X(T)  = \frac{1}{\sqrt{6}}\left( T_{F1}k +
\frac{16(T_{F1}k\pi)^{3/2}T^{3}}{81T_{D3}}+
\frac{40T_{F1}^{2}k^{2}\pi^{3}T^{6}}{729T_{D3}^{2}}\right)^{1/2}.\label{Q10}
\end{eqnarray}
At the beginning, we have  $T = \infty$ that decreases with time. On
the other hand, Eq. (\ref{Q10}) shows that,  at this time, the
scale factor is zero and  with the decreasing of temperature, the Universe
expands.

After a short period, the wormhole is  formed between brane
and antibrane due to the $F$-string charge and  the Universe  is  entering  the
inflationary phase. Assuming  $k$ units for the  $F$-string charge along the
radial direction and using Eq. (\ref{Q5}), we obtain
\cite{q13,q14}:
\begin{eqnarray}
z(\sigma)= \int_{\sigma}^{\infty}
d\acute{\sigma}(\frac{F(\acute{\sigma})^{2}}{F(\sigma_{0})^{2}}-1)^{-\frac{1}{2}}.
\label{Q11}
\end{eqnarray}
At  finite temperature BIon configuration, the  $F(\sigma)$ is given by
\begin{eqnarray}
F(\sigma) = \sigma^{2}\frac{4\cosh^{2}\alpha - 3}{\cosh^{4}\alpha},
\label{Q12}
\end{eqnarray}
where $\cosh{\alpha}$ is determined by the following function:
\begin{eqnarray}
\cosh^{2}\alpha = \frac{3}{2}\frac{\cos\frac{\delta}{3} +
\sqrt{3}\sin\frac{\delta}{3}}{\cos\delta}, \label{Q13}
\end{eqnarray}
with the definitions:
\begin{eqnarray}
\cos{\delta} \equiv \overline{T}^{4}\sqrt{1 +
\frac{k^{2}}{\sigma^{4}}},\quad  \overline{T} \equiv
\left(\frac{9\pi^{2}N}{4\sqrt{3}T_{D_{3}}}\right)T, \quad \kappa \equiv \frac{k
T_{F1}}{4\pi T_{D_{3}}} \label{Q14}
\end{eqnarray}
In the last  equation, $T$ is the finite temperature of the BIon system, $N$ is the
number of $D3$-branes,  $T_{D_{3}}$ and $T_{F1}$ are the tensions of
branes and fundamental strings respectively. Attaching a mirror
solution to Eq. (\ref{Q11}), we construct the wormhole configuration.
The estimation of separation distance $\Delta = 2z(\sigma_{0})$
between the $N$ $D3$-branes and $N$ anti-$D3$-branes for a given
brane-antibrane wormhole configuration depends on the four
parameters $N$, $k$, $T$ and $\sigma_{0}$. We have:
\begin{eqnarray}
\Delta = 2z(\sigma_{0})= 2\int_{\sigma_{0}}^{\infty}
d\acute{\sigma}\left(\frac{F(\acute{\sigma})^{2}}{F(\sigma_{0})^{2}}-1\right)^{-\frac{1}{2}}.
\label{Q15}
\end{eqnarray}
In in the limit of small temperatures, we obtain:
\begin{eqnarray}
\Delta = \frac{2\sqrt{\pi}\Gamma(5/4)}{\Gamma(3/4)}\sigma_{0}\left(1 +
\frac{8}{27}\frac{k^{2}}{\sigma_{0}^{4}}\overline{T}^{8}\right).
\label{Q16}
\end{eqnarray}
Let us now discuss the non-phantom  inflationary model of universe in the  thermal BIon system.
In order to discuss this scenario, we  have to compute the contribution of the BIonic system
to the four-dimensional  energy-momentum tensor. The energy-momentum  tensor for a
BIonic system with $N$ $D3$-branes and $k$ $F$-string charges is \cite{q14},
 \begin{eqnarray}
&& T^{00}=\frac{2T_{D3}^{2}}{\pi
T^{4}}\frac{F(\sigma)}{\sqrt{F^{2}(\sigma)-F^{2}(\sigma_{0})}}\sigma^{2}\frac{4\cosh^{2}\alpha
+ 1}{\cosh^{4}\alpha} \nonumber \\&& T^{ii}=
-\gamma^{ii}\frac{8T_{D3}^{2}}{\pi
T^{4}}\frac{F(\sigma)}{\sqrt{F^{2}(\sigma)-F^{2}(\sigma_{0})}}\sigma^{2}\frac{1}{\cosh^{2}\alpha},\,i=1,2,3
\nonumber \\&&T^{44}=\frac{2T_{D3}^{2}}{\pi
T^{4}}\frac{F(\sigma)}{F(\sigma_{0})}\sigma^{2}\frac{4\cosh^{2}\alpha
+ 1}{\cosh^{4}\alpha} \label{Q17}
\end{eqnarray}
We assume this higher-dimensional stress-energy tensor  to be
a  perfect fluid   of the form  ($ T_i^j = {\mathop{\rm
diag}\nolimits} \left[ {-\rho,  p,  p,  p,  \bar{p},  p,  p,
 p} \right])$ where $\bar{p}$ is the pressure in the extra
space-like dimension. In above the equation, we   allow  the pressure
in the extra dimension to be different with respect to the
pressure in the $3D$ space. Therefore, this stress-energy tensor
expresses a homogeneous, anisotropic perfect fluid in ten
dimensions. This equation shows that with increasing temperature
in BIonic system, the energy-momentum tensors decreases. This is
because that when spikes of  branes and antibranes are well
separated, wormhole is not formed and there is no channel
for flowing energy from universe branes into extra dimensions. This means that
temperature is very high.  However when the  two universe branes are
close to each other and connected by a wormhole, temperature
reduces to lower values.

Now, we can discuss the  phantom cosmological model in finite
temperature BIon configuration  and obtain the explicit form of temperature and
equation of state parameter $\omega$. To this end, we use  the
approach reported in Ref. \cite{q12} in order to unify BIonic and phantom inflation through
the  three phases of universe expansion.

A phantom cosmological
model can be described by  the following action:
\begin{eqnarray}
&& S=\int d^{4}x \sqrt{-g}\{\frac{1}{2k^{2}}R -
\frac{1}{2}\omega(\phi)\partial_{\mu}\phi
\partial^{\mu}\phi - V(\phi)\}\label{Q18}
\end{eqnarray}
Here, $\omega(\phi)$ and $V(\phi)$ are functions of the scalar
field $\phi$. The energy density $\rho$ and the pressure $p$
are:
\begin{eqnarray}
&& \rho = \frac{1}{2}\omega(\phi)\dot{\phi}^{2} +
V(\phi),\nonumber \\&& p = \frac{1}{2}\omega(\phi)\dot{\phi}^{2} -
V(\phi). \label{Q19}
\end{eqnarray}
Furthermore, the FRW cosmological equations are given by \cite{q12}:
\begin{eqnarray}
&& \rho_{uni1} = \rho_{uni2} = \frac{3}{k^{2}}H^{2},\nonumber
\\&& p_{uni1} = p_{uni2}=-\frac{1}{k^{2}}(3H^{2} + 2\dot{H})\nonumber
\\&& \rho_{tot}=2\rho_{uni1}, p_{tot}=2 p_{uni1}.
\label{Q20}
\end{eqnarray}
Using these FRW equations, the effective equation of state is:
\begin{eqnarray}
&& \omega_{eff} = \frac{p_{tot}}{\rho_{tot}} = -1 -
\frac{2}{3}\frac{\dot{H}}{H^{2}}. \label{Q21}
\end{eqnarray}
Now, the scalar field $\phi$, the Hubble rate H and the
scale factor $a(t)$ can be chosen follow as:
\begin{eqnarray}
&& \phi = t,\nonumber
\\&& H = h_{0}^{2}\left(\frac{1}{t_{0}^{2} - \phi^{2}} + \frac{1}{t_{1}^{2} + \phi^{2}}\right),\nonumber
\\&&
a(t)=a_{0}\left(\frac{t+t_{0}}{t_{0}-t}\right)^{\frac{h_{0}^{2}}{2t_{0}}}e^{-\frac{h_{0}^{2}}{2t_{0}}Arctan(\frac{t_{1}}{t})}.
\label{Q22}
\end{eqnarray}
Then, using Eqs. (\ref{Q21}) and (\ref{Q22}), the effective
EoS parameter is written as \cite{q11, q12}:
\begin{eqnarray}
&& \omega_{eff} =\frac{p}{\rho} = -1
-\frac{8}{3h_{0}^{2}}\frac{t(t-t_{+})(t-t_{-})}{(t_{1}^{2}+t_{0}^{2})^{2}}\label{Q23}
\end{eqnarray}
Since a = 0 at $t =- t_{0}$, one may regard this time corresponding
to the birth of the universe. We find that $H$ has two minima  at $t =
t_{\pm} = \pm \sqrt{\frac{t_{0}^{2}-t_{1}^{2}}{2}}$ and at $t = 0$. Besides
$H$ has a local maximum. Hence, the phantom phase ($\omega_{eff} < -1$)
occurs for  $t_{-}< t < 0$ and $t
> t_{+}$, while the  non-phantom phase ($\omega_{eff} > -1$) for $-t_{0}< t <t_{-}$ and $0 < t < t_{+}$. It is worth noticing
 that there is a Big Rip type singularity at t = $t_{0}$ \cite{q11, q12}.

Now, using Eq. (\ref{Q17}), we  obtain the equation of
state on the universe brane in the finite temperature BIon configuration:
\begin{eqnarray}
&& \omega_{BIon} = -\frac{4 \cosh^{2}\alpha}{4\cosh^{2}\alpha +1
}\left(1+\frac{(t_{-})^{2}-(t-t_{-})^{2}}{(t-t_{-})^{2}}\right).\label{Q24}
\end{eqnarray}
As it can been seen from Eq. (\ref{Q24}), the equation of state
is  less than -1 in the range of $t_{-}<t<0$
and it is evaluated from phantom  to non-phantom phase at $t=0$. Equating
 this equation of state  with equation of state  in Eq.
(\ref{Q23}) , we can find the explicit form of temperature $T$,  that is
\begin{eqnarray}
&& T \sim
\left(1+\frac{(t_{-})^{2}-(t-t_{-})^{2}}{(t-t_{-})^{2}}\right)^{1/4}\frac{1}{\left(1+
\frac{8}{3h_{0}^{2}}\frac{t(t-t_{+})(t-t_{-})}{(t_{1}^{2}+t_{0}^{2})^{2}}\right)}.\label{Q25}
\end{eqnarray}
Eq. (\ref{Q25}) indicates that temperature is infinite  at
$t=t_{-}$ and decreases with time. However, the velocity of this
decreasing is very high in the range of $t_{-}<t<0$.  This result is in good agreement with observational data.

We assume that the wormhole is created at $t = t_{-}$ and
$\sigma=\sigma_{0}$ and it  vanishes  at $t = 0$ and
$\sigma_{0}=0$. In this period of time, we can write: $\sigma_{0}
= \frac{0 -t}{0 -t_{-}}\sigma$. Using this and putting the energy
density of the  two universes equal to the energy density of the  BIon, we
obtain $\sigma$ in terms of time:
\begin{eqnarray}
&& \rho_{tot} = \rho_{BIon} \rightarrow
\frac{6}{k^{2}}H^{2}=\frac{2T_{D3}^{2}}{\pi
T^{4}}\frac{F(\sigma)}{\sqrt{F^{2}(\sigma)-F^{2}(\sigma_{0})}}\sigma^{2}\frac{4\cosh^{2}\alpha
+ 1}{\cosh^{4}\alpha}\rightarrow \nonumber \\&& \sigma \sim \left(1+\frac{(t_{-})^{2}-(t-t_{-})^{2}}{(t-t_{-})^{2}}
\right)^{-1/4}\left(\frac{(t-t_{-})}{(t_{-})^{2}-(t-t_{-})^{2}}\right)\left(1+
\frac{8}{3h_{0}^{2}}\frac{t(t-t_{+})(t-t_{-})}{(t_{1}^{2}+t_{0}^{2})^{2}}\right)
\label{Q26}
\end{eqnarray}
According to this result, $\sigma$ is zero at $t = t_{-}$;
however, with  time evolution, it accelerates and tends to very higher
values in a short period. From this point of view, the
behavior of  $\sigma$ is the same as the scale factor $a(t)$.

\section{Stage 2: The  non-phantom standard
cosmology }\label{o2} In this section, we propose a model that
allows to consider the non-phantom model in the brane-antibrane
system. In this stage, with decreasing temperature and distance
between two branes, the wormhole between brane and anti-brane evaporates and   tachyon is born.  The expansion of the two FRW universes
is controlled by the tachyonic potential between branes and
evolves from non-phantom to phantom phase.

To construct  a non-phantom model, we consider a set of
$D3$-$\overline{D3}$-brane pairs in the background  (\ref{Q5}) which
are placed at points $z_{1} = l/2$ and $z_{2} = -l/2$ respectively
so that the separation between the brane and antibrane is $l$. For
the simple case of a single $D3$-$\overline{D3}$-brane pair with
open string tachyon, the action is \cite{q16}:
 \begin{eqnarray}
&& S=-\tau_{3}\int d^{9}\sigma \sum_{i=1}^{2}
V(TA,l)e^{-\phi}(\sqrt{-det A_{i}})\nonumber \\&&
(A_{i})_{ab}=\left(g_{MN}-\frac{TA^{2}l^{2}}{Q}g_{Mz}g_{zN}\right)\partial_{a}x^{M}_{i}\partial_{b}x^{M}_{i}
+F^{i}_{ab}+\frac{1}{2Q}((D_{a}TA)(D_{b}TA)^{\ast}+(D_{a}TA)^{\ast}(D_{b}TA))\nonumber
\\&&
+il(g_{az}+\partial_{a}z_{i}g_{zz})(TA(D_{b}TA)^{\ast}-TA^{\ast}(D_{b}TA))+
il(TA(D_{a}TA)^{\ast}-TA^{\ast}(D_{a}TA))(g_{bz}+\partial_{b}z_{i}g_{zz}),
\label{Q27}
\end{eqnarray}
where
  \begin{eqnarray}
&& Q=1+TA^{2}l^{2}g_{zz}, \nonumber \\&&
D_{a}TA=\partial_{a}TA-i(A_{2,a}-A_{1,a})TA,
V(TA,l)=g_{s}V(TA)\sqrt{Q}, \nonumber \\&& e^{\phi}=g_{s}( 1 +
\frac{R^{4}}{z^{4}} )^{-\frac{1}{2}}, \label{Q28}
\end{eqnarray}
The quantities $\phi$, $A_{2,a}$ and $F^{i}_{ab}$ are the dilaton field, the gauge
fields and field strengths on the world-volume of the non-BPS
brane respectively;  $TA$ is the tachyon field, $\tau_{3}$ is the brane tension
and $V(TA)$ is the tachyon potential. The indices $a,b$ denote the
tangent directions of $D$-branes, while the indices $M,N$ run over the
background ten-dimensional space-time directions. The $Dp$-brane and
the anti-$Dp$-brane are labeled by $i$ = 1 and 2 respectively. Then
the separation between these $D$-branes is defined by $z_{2} - z_{1}
= l$. Also, in writing the above action, we are using the convention
$2\pi\acute{\alpha}=1$.

Let us consider, for simplicity,  the only $\sigma$
dependence of the tachyon field $TA$  and set the
gauge fields to zero. In this case, the action (\ref{Q27}) in the
region that  $r> R$ and $TA'\sim constant$ simplifies to
  \begin{eqnarray}
S \simeq-\frac{\tau_{3}}{g_{s}}\int dt \int d\sigma \sigma^{2}
V(TA)(\sqrt{D_{1,TA}}+\sqrt{D_{2,TA}}), \label{Q29}
\end{eqnarray}
where $D_{1,TA} = D_{2,TA}\equiv D_{TA}$,
${\displaystyle V_{3}=\frac{4\pi^{2}}{3}}$ is the volume of a unit sphere $S^{3}$ and
 \begin{eqnarray}
D_{TA} = 1 + \frac{l'(\sigma)^{2}}{4}+ TA^{2}l^{2}, \label{Q30}
\end{eqnarray}
where the prime  denotes a derivative with respect to
$\sigma$. A useful potential  that can be used is
\cite{q17,q18,q19}:
 \begin{eqnarray}
V(TA)=\frac{\tau_{3}}{\cosh\sqrt{\pi}TA}. \label{Q31}
\end{eqnarray}
The energy momentum tensor is obtained from the action by calculating
its functional derivative with respect to the  ten-dimensional background
metric $g_{MN}$. The  variation  is ${\displaystyle T^{MN} =
\frac{2}{\sqrt{-det g}}\frac{\delta S}{\delta g_{MN}}}$. We get
\cite{q16},
 \begin{eqnarray}
&& T^{00}_{i}=V(TA)\sqrt{D_{TA}},  \nonumber \\&&
T^{zz}_{i}=-V(TA)\frac{1}{\sqrt{D_{TA}}}
(TA^{2}l^{2}+\frac{\acute{l}^{2}}{4}) \nonumber \\&&
T^{\sigma\sigma}_{i} = -V(TA)\frac{Q}{\sqrt{D_{TA}}} \label{Q32}
\end{eqnarray}
Now, using the above equation, we  obtain the equation of state as:
\begin{eqnarray}
&& \omega_{brane-antibrane} = -\frac{1+TA^{2}l^{2}}{1 +
\frac{l'^{2}}{4}+ TA^{2}l^{2}}\label{Q33}
\end{eqnarray}
This equation indicates that the equation of state is
negative  both at the  beginning and at the end of this era and bigger than
-1 in the range of $0<t<t_{+}$. Assuming the equation of state  equal
to the equation of state  in (\ref{Q23}) (which corresponds to the unified
theory and can be applied for all the three phases) and assuming
$\sigma\sim t$, $l \sim l_{0}(1
-\frac{t_{+}t^{2}}{2}+\frac{t^{3}}{3})$ and $l' \sim
l_{0}t(t-t_{+})$ , we get:
\begin{eqnarray}
&& TA \sim \frac{t^{4}}{(t_{1}^{2}+t_{0}^{2})\left(2
+\frac{8}{3h_{0}^{2}}\frac{t(t_{+}-t)(t-t_{-})}{(t_{1}^{2}+t_{0}^{2})^{2}}\right)}
\label{Q34}
\end{eqnarray}
Eq. (\ref{Q34} ) shows that when two branes are very
distant  from each other (t=0, $l=l_{0}$), the tachyon field is zero ,
whereas moving the branes towards each other, the value of tachyon
increases and becomes very large at $t = t_{+}$.

\section{ Stage 3: The  late-time
acceleration }\label{o3}

In the previous section, we considered  that the tachyon field
grows slowly ($TA \sim t^{4}/t^{3}= t$) and we ignored
${\displaystyle TA'=\frac{\partial TA}{\partial \sigma}}$ and
${\displaystyle \dot{TA}=\frac{\partial TA}{\partial t}}$ in our calculations. In
this section, we discuss that with the decreasing of the distance separation
 between the brane and antibrane universes, the tachyon field grows very fast and
$TA'$ and $\dot{TA}$ cannot be discarded. This dynamics leads to   the formation of a  new wormhole.
 In this stage, the Universe evolves from non-phantom phase to a new
phantom phase and consequently, the phantom-dominated era of the
universe accelerates and ends up into the  Big-Rip singularity. In this
case, the action (\ref{Q27}) is given by the following Lagrangian $L$:
  \begin{eqnarray}
L \simeq-\frac{\tau_{3}}{g_{s}} \int d\sigma \sigma^{2}
V(TA)(\sqrt{D_{1,TA}}+\sqrt{D_{2,TA}}), \label{Q35}
\end{eqnarray}
where
 \begin{eqnarray}
D_{1,TA} = D_{2,TA}\equiv D_{TA} = 1 + \frac{l'(\sigma)^{2}}{4}+
\dot{TA}^{2} -  TA'^{2}, \label{Q36}
\end{eqnarray}
where we assume that $TA l\ll TA'$. Now, we  study the Hamiltonian
corresponding to the above Lagrangian.
 In order to derive such Hamiltonian,  we need the canonical momentum density ${\displaystyle \Pi =
\frac{\partial L}{\partial \dot{TA}}}$ associated with the tachyon, that is
 \begin{eqnarray}
\Pi = \frac{V(TA)\dot{TA}}{ \sqrt{1 + \frac{l'(\sigma)^{2}}{4}+
\dot{TA}^{2} -  TA'^{2}}}, \label{Q37}
\end{eqnarray}
so that the Hamiltonian can be obtained as:
\begin{eqnarray}
H_{DBI} = 4\pi\int d\sigma  \sigma^{2} \Pi \dot{TA} - L.
 \label{Q38}
\end{eqnarray}
By choosing $\dot{TA} = 2 TA'$, this gives:
\begin{eqnarray}
H_{DBI} = 4\pi\int d\sigma \sigma^{2} \left[\Pi
(\dot{TA}-\frac{1}{2}TA')\right] + \frac{1}{2}TA\partial_{\sigma}(\Pi
\sigma^{2}) - L
 \label{Q39}
\end{eqnarray}
In this equation, we have, in the second step, integrated by parts
the term proportional to $\dot{TA}$, indicating that  tachyon can
be studied as a Lagrange multiplier imposing the constraint
$\partial_{\sigma}(\Pi \sigma^{2}V(TA))=0$ on the canonical
momentum. Solving this equation yields:
\begin{eqnarray}
\Pi =\frac{\beta}{4\pi \sigma^{2}},
 \label{Q40}
\end{eqnarray}
where $\beta$ is a constant. Using (\ref{Q40}) in (\ref{Q38}), we
get:
\begin{eqnarray}
&& H_{DBI} = \int d\sigma V(TA)\sqrt{1 + \frac{l'(\sigma)^{2}}{4}
+ \dot{TA}^{2} -  TA'^{2}}F_{DBI} ,  \nonumber \\&&
F_{DBI}=\sigma^{2}\sqrt{1 + \frac{\beta}{\sigma^{2}}}\label{Q41}
\end{eqnarray}
The resulting equation of motion  for $l(\sigma)$, calculating by varying
(\ref{Q41}), is
\begin{eqnarray}
&&\left(\frac{l'F_{DBI}}{4\sqrt{1+
\frac{l'(\sigma)^{2}}{4}}}\right)'=0\label{Q42}
\end{eqnarray}
Solving this equation, we obtain:
\begin{eqnarray}
&&l(\sigma) = 4\int_{\sigma}^{\infty} d\sigma
\left(\frac{F_{DBI}(\sigma)}{F_{DBI}(\sigma_{0})}-1\right)^{-\frac{1}{2}}=4\int_{\sigma}^{\infty}
d\sigma'\left(\frac{\sqrt{\sigma_{0}^{4}+\beta^{2}}}{\sqrt{\sigma'^{4}-\sigma_{0}^{4}}}\right)
\label{Q43}
\end{eqnarray}
This solution, for non-zero $\sigma_{0}$,   represents a wormhole
with a finite size throat. However, this solution is not complete,
because  we ignored the acceleration of branes. This acceleration
is due to the tachyon potential between the branes ( ${\displaystyle a\sim \frac{\partial
V(T)}{\partial \sigma}}$). According to recent investigations
\cite{q20}, each of the accelerated branes and antibranes detects the
Unruh temperature(${\displaystyle T=\frac{\hbar a}{2k_{B}\pi c}}$). We will show
that this system is equivalent to the black brane.
The equation of motion obtained from action (\ref{Q41})  is:
\begin{eqnarray}
\left(\frac{1}{\sqrt{D_{TA}}}TA'(\sigma)\right)'=\frac{1}{\sqrt{D_{TA}}}
\left[\frac{V'(TA)}{V(TA)}(D_{TA}-TA'(\sigma)^{2})\right] \label{Q44}
\end{eqnarray}
We can reobtain this equation in accelerated fame from the
equation of motion in the flat background of (\ref{Q5}):
\begin{eqnarray}
&& -\frac{\partial^{2} TA}{\partial \tau^{2}} + \frac{\partial^{2}
TA}{\partial \sigma^{2}}=0 \label{Q45}
\end{eqnarray}
By using the following re-parameterizations
\begin{eqnarray}
&& \rho = \frac{\sigma^{2}}{w} ,  \nonumber \\&& w=
\frac{V(TA)\sqrt{D_{TA}}F_{DBI}}{2M_{D3-brane}}\nonumber \\&&
\bar{\tau} = \gamma\int_{0}^{t} d\tau' \frac{w}{\dot{w}} - \gamma
\frac{\sigma^{2}}{2}\label{Q46}
\end{eqnarray}
and doing following calculations:
\begin{eqnarray}
\left\{\left[\left(\frac{\partial \bar{\tau}}{\partial \tau}\right)^{2} -
\left(\frac{\partial \bar{\tau}}{\partial
\sigma}\right)^{2}\right]\frac{\partial^{2}}{\partial \tau^{2}}+
\left[\left(\frac{\partial \rho}{\partial \sigma}\right)^{2} - \left(\frac{\partial
\rho}{\partial \tau}\right)^{2}\right]\frac{\partial^{2}}{\partial
\rho^{2}}\right\}TA=0\label{Q47}
\end{eqnarray}
we have:
\begin{eqnarray}
&& (-g)^{-1/2}\frac{\partial}{\partial
x_{\mu}}\left[(-g)^{1/2}g^{\mu\nu}\right]\frac{\partial}{\partial
x_{\upsilon}}TA=0\label{Q48}
\end{eqnarray}
where $x_{0}=\bar{\tau}$, $x_{1}=\rho$ and the metric elements are
obtained as:
\begin{eqnarray}
&& g^{\bar{\tau}\bar{\tau}}\sim
-\frac{1}{\beta^{2}}\left(\frac{w'}{w}\right)^{2}\frac{\left(1-\left(\frac{w}{w'}\right)^{2}\frac{1}{\sigma^{4}}\right)}
{\left(1+\left(\frac{w}{w'}\right)^{2}\frac{(1+\gamma^{-2})}{\sigma^{4}}\right)^{1/2}}\nonumber
\\&&g^{\rho\rho}\sim -(g^{\bar{\tau}\bar{\tau}})^{-1}\label{Q49}
\end{eqnarray}
where we have used of previous assumption($\frac{\partial
TA}{\partial t} = \frac{\partial TA}{\partial \tau}= 2
\frac{\partial TA}{\partial \sigma}$).

Now, we can compare these elements with the line elements of one
black $D3$-brane \cite{q21}:
\begin{eqnarray}
&& ds^{2}= D^{-1/2}\bar{H}^{-1/2}(-f
dt^{2}+dx_{1}^{2})+D^{1/2}\bar{H}^{-1/2}(dx_{2}^{2}+dx_{3}^{2})+D^{-1/2}\bar{H}^{1/2}(f^{-1}
dr^{2}+r^{2}d\Omega_{5})^{2},\nonumber
\\&&\label{Q50}
\end{eqnarray}
where
\begin{eqnarray}
&&f=1-\frac{r_{0}^{4}}{r^{4}},\nonumber
\\&&\bar{H}=1+\frac{r_{0}^{4}}{r^{4}}\sinh^{2}\alpha , \nonumber
\\&&D^{-1}=\cos^{2}\varepsilon + H^{-1}\sin^{2}\varepsilon, \nonumber
\\&&\cos\varepsilon =\frac{1}{\sqrt{1+\frac{\beta^{2}}{\sigma^{4}}}}.
\label{Q51}
\end{eqnarray}
Eqs. (\ref{Q49}) and (\ref{Q50}) lead to
\begin{eqnarray}
&&f=1-\frac{r_{0}^{4}}{r^{4}}\sim
1-\left(\frac{w}{w'}\right)^{2}\frac{1}{\sigma^{4}},\nonumber
\\&&\bar{H}=1+\frac{r_{0}^{4}}{r^{4}}\sinh^{2}\alpha \sim 1+\left(\frac{w}{w'}\right)^{2}\frac{(1+\gamma^{-2})}{\sigma^{4}} \nonumber
\\&&D^{-1}=\cos^{2}\varepsilon + \bar{H}^{-1}sin^{2}\varepsilon\simeq1\nonumber
\\&&
\Rightarrow r\sim \sigma,r_{0}\sim
\left(\frac{w}{w'}\right)^{1/2},(1+\gamma^{-2})\sim \sinh^{2}\alpha
\label{Q52}
\end{eqnarray}
The temperature of the BIon system  is ${\displaystyle T=\frac{1}{\pi r_{0} \cosh\alpha}}$
\cite{q13}. Consequently, the temperature of the brane-antibrane
system can be calculated as:
\begin{eqnarray}
&& T=\frac{1}{\pi r_{0}
\cosh\alpha}=\frac{\gamma}{\pi}\left(\frac{w'}{w}\right)^{1/2}\sim \nonumber
\\&&\frac{\gamma}{\pi}\left(\tanh\sqrt{\pi}TA+\frac{l'l''+TA'TA''}{1 + \frac{l'(\sigma)^{2}}{4}+
 TA'^{2}}+ \frac{\frac{\beta}{\sigma^{3}}}{1+\frac{\beta}{\sigma^{2}}}\right) \label{Q53}
\end{eqnarray}
However, this result should be corrected. Because $\gamma$ depends
on the temperature and we can write:
\begin{eqnarray}
&&\gamma = \frac{1}{\cosh\alpha} \sim
\frac{2\cos\delta}{3\sqrt{3}-\cos\delta
-\frac{\sqrt{3}}{6}\cos^{2}\delta}\sim \nonumber
\\&&\frac{2\bar{T}^{4}\sqrt{1+\frac{\beta^{2}}{\sigma^{4}}}}{3\sqrt{3}-\bar{T}^{4}\sqrt{1+\frac{\beta^{2}}{\sigma^{4}}}
-\frac{\sqrt{3}}{6}\bar{T}^{8}(1+\frac{\beta^{2}}{\sigma^{4}})}
\label{Q54}
\end{eqnarray}
Using Eqs. (\ref{Q53}) and (\ref{Q54}), we can approximate
the explicit form of temperature:
\begin{eqnarray}
&&T \sim
\left(\frac{4\sqrt{3}T_{D_{3}}}{9\pi^{2}N}\right)\frac{\sqrt[3]{\pi}}{\sqrt[6]{1+\frac{\beta^{2}}{\sigma^{4}}}}
\left(\tanh\sqrt{\pi}TA+\frac{l'l''+TA'TA''}{1 +
\frac{l'(\sigma)^{2}}{4}+
 TA'^{2}}+ \frac{\frac{\beta}{\sigma^{3}}}{1+\frac{\beta}{\sigma^{2}}}\right)^{-1/3}
\label{Q55}
\end{eqnarray}
This equation shows that with approaching the two branes together and
increasing the tachyon, the temperature of system decreases. This result
is consistent with the thermal history of universe that temperature
decreases with time. Now, we want to estimate the dependency of the
tachyon on time. To this end, we calculate the energy momentum
tensor components and equation of state. Using the energy-momentum tensor for the
black $D3$-brane\cite{q13}, we obtain:
\begin{eqnarray}
&& T^{00} =
\frac{\pi^{2}}{2}T_{D3}^{2}r_{0}^{4}(5+4\sinh^{2}\alpha)\sim
\frac{\pi^{2}}{2}T_{D3}^{2}\left(\frac{w}{w'}\right)^{1/2}(9 +
\gamma^{-2})\sim \nonumber
\\&& \frac{\pi^{2}}{2}T_{D3}^{2}\left(\tanh\sqrt{\pi}TA+\frac{l'l''+TA'TA''}{1 + \frac{l'(\sigma)^{2}}{4}+
 TA'^{2}}+\frac{\frac{\beta}{\sigma^{3}}}{1+\frac{\beta}{\sigma^{2}}}\right)^{-1}\left(9+\frac{2\bar{T}^{4}
 \sqrt{1+\frac{\beta^{2}}{\sigma^{4}}}}{3\sqrt{3}-\bar{T}^{4}\sqrt{1+\frac{\beta^{2}}{\sigma^{4}}}
-\frac{\sqrt{3}}{6}\bar{T}^{8}(1+\frac{\beta^{2}}{\sigma^{4}})}\right)
\nonumber
\\&& \nonumber
\\&& \nonumber
\\&& T^{ii} =-\gamma^{ii}
\frac{\pi^{2}}{2}T_{D3}^{2}r_{0}^{4}(1+4\sinh^{2}\alpha)\sim
-\left(1+\frac{l'^{2}}{4}\right)\frac{\pi^{2}}{2}T_{D3}^{2}\left(\frac{w}{w'}\right)^{1/2}(5
+ \gamma^{-2})\sim \nonumber
\\&& -\left(1+\frac{l'^{2}}{4}\right)\frac{\pi^{2}}{2}T_{D3}^{2}\left(\tanh\sqrt{\pi}TA+\frac{l'l''+TA'TA''}
{1 + \frac{l'(\sigma)^{2}}{4}+ TA'^{2}}+
 \frac{\frac{\beta}{\sigma^{3}}}{1+\frac{\beta}{\sigma^{2}}}\right)^{-1}\times \nonumber
\\&&\left(5+\frac{2\bar{T}^{4}\sqrt{1+\frac{\beta^{2}}{\sigma^{4}}}}{3\sqrt{3}-\bar{T}^{4}\sqrt{1+\frac{\beta^{2}}{\sigma^{4}}}
-\frac{\sqrt{3}}{6}\bar{T}^{8}(1+\frac{\beta^{2}}{\sigma^{4}})}\right)
\label{Q56}
\end{eqnarray}
We assume that the wormhole was created at $t = t_{+}$ and
$\sigma=\sigma_{0}$ and will be vanished at $t = t_{rip}$ and
$\sigma_{0}=0$. In this period of time, we can write: $\sigma_{0}
= \frac{ t -t_{+}}{ t_{rip} -t_{+}}\sigma$. Using this and the
relation( $ T_i^j = {\mathop{\rm diag}\nolimits} \left[ {\rho, -
p, - p, - p, - \bar{p}, - p, - p, - p,
 } \right])$, we can calculate the equation of state parameter:
\begin{eqnarray}
&& \omega_{BIon} = -\frac{(t_{rip}-t_{+})^{2}(1+\beta^{2}
+(t-t_{+})^{2})\left(5+\frac{2\bar{T}^{4}\sqrt{1+\frac{\beta^{2}}{\sigma^{4}}}}{3\sqrt{3}-\bar{T}^{4}\sqrt{1+\frac{\beta^{2}}
{\sigma^{4}}} -\frac{\sqrt{3}}{6}\bar{T}^{8}(1+\frac{\beta^{2}}{\sigma^{4}})}\right)}{(t_{rip}-t)(t_{rip}-t
+2t_{+})\left(9+\frac{2\bar{T}^{4}\sqrt{1+\frac{\beta^{2}}{\sigma^{4}}}}{3\sqrt{3}-\bar{T}^{4}\sqrt{1+\frac{\beta^{2}}{\sigma^{4}}}
-\frac{\sqrt{3}}{6}\bar{T}^{8}(1+\frac{\beta^{2}}{\sigma^{4}})}\right)}\label{Q57}
\end{eqnarray}
For $\beta>\frac{2}{\sqrt{5}}$, the equation of state parameter is
negative one at the beginning  of this era and less than -1 in the
range of $t_{+}<t<t_{rip}$. Putting this EOS parameter equal to
EOS parameter in (\ref{Q23}) (which corresponds to unified theory
and can be applied for all three phases), we get:
\begin{eqnarray}
&& T\sim \frac{(t_{rip}-t)(t_{rip}-t
+2t_{+})}{(t_{rip}-t_{+})^{2}(1+\beta^{2} +(t-t_{+})^{2})\left(1
+\frac{8}{3h_{0}^{2}}\frac{t(t-t_{+})(t-t_{-})}{(t_{1}^{2}+t_{0}^{2})^{2}}\right)\left(\frac{-(t_{rip}-t)(t_{rip}-t
+2t_{+})}{(t_{rip}-t_{+})^{2}(1+\beta^{2}
+(t-t_{+})^{2})}+1\right)}\label{Q58}
\end{eqnarray}
This equation shows that temperature decreases with time and tends
to zero at Big Rip singularity. As can be seen from temperatures
in three stages of universe, temperature was infinite  at the
beginning, reduces very fast in the inflation era, decreases with
lower velocity in the non-phantom phase,  and finally reduces with higher rate
at the late-time acceleration  converging  to zero at the ripping
time. This result is in agreement with recent observations and also
with thermal history of universe.

\section{ Testing the model against observational data}\label{o4}
In previous sections, we proposed an approach to  unify inflation, deceleration and
acceleration phases of the Universe. In this section, we compare qualitatively  the
model with cosmological data and obtain some  results
like the ripping time. To this end, we calculate the
deceleration parameter in each era of expansion history. It is
\begin{eqnarray}
&& q =- \frac{1}{H^{2}}\frac{dH}{dt}-1\label{Q59}
\end{eqnarray}
Using the relation
$6H^{2}=\rho_{Uni1}+\rho_{Uni2}=\rho_{brane-antibrane}$ and
Eqs. (\ref{Q26}),(\ref{Q35}) and (\ref{Q56}), we  find the
deceleration parameter in the three stages:
\begin{eqnarray}
\begin{array}{cc}
q\sim-(\frac{(t_{-})^{2}-(t-t_{-})^{2}}{(t-t_{-})^{2}})^{4}
\Bigl[\frac{8[(t-t_{+})(t-t_{-})+t(t-t_{-})+t(t-t_{+})]}{3h_{0}^{2}+
8\frac{t(t-t_{+})(t-t_{-})}{(t_{1}^{2}+t_{0}^{2})^{2}}} \\ +
2\frac{((t_{-})^{2}-(t-t_{-})^{2})(t-t_{-})+(t-t_{-})^{3}}{(1+\frac{(t_{-})^{2}-(t-t_{-})^{2}}{(t-t_{-})^{2}})^{3/2}}
+2\frac{(t-t_{-})^{3}+(t-t_{-})((t_{-})^{2}-(t-t_{-})^{2})}{((t_{-})^{2}-(t-t_{-})^{2})^{2}}\Bigr]
& t_{-}<t<0 \\ \\
q\sim \tanh\left(\sqrt{\pi}\frac{t^{4}(t_{+}-t)}{(t_{1}^{2}+t_{0}^{2})(2
+\frac{8}{3h_{0}^{2}}\frac{t(t_{+}-t)(t-t_{-})}{(t_{1}^{2}+t_{0}^{2})^{2}})}\right)\Bigl[1+\frac{t^{4}(t_{+}-t)}
{(t_{1}^{2}+t_{0}^{2})(2 +\frac{8}{3h_{0}^{2}}\frac{t(t_{+}-t)(t-t_{-})}{(t_{1}^{2}+t_{0}^{2})^{2}})}
+ t(t_{+}-t)\Bigr]^{1/2}\\+
sinh^{2}\left(\sqrt{\pi}\frac{t^{4}(t_{+}-t)}{(t_{1}^{2}+t_{0}^{2})(2
+\frac{8}{3h_{0}^{2}}\frac{t(t_{+}-t)(t-t_{-})}{(t_{1}^{2}+t_{0}^{2})^{2}})}\right)
\frac{(t_{+}-2t)+\frac{t^{3}(4t_{+}-5t)}{(t_{1}^{2}+t_{0}^{2})(2
+\frac{8}{3h_{0}^{2}}\frac{t(t_{+}-t)(t-t_{-})}{(t_{1}^{2}+t_{0}^{2})^{2}})}}{\left[1+\frac{t^{4}(t_{+}-t)}{(t_{1}^{2}+t_{0}^{2})(2
+\frac{8}{3h_{0}^{2}}\frac{t(t_{+}-t)(t-t_{-})}{(t_{1}^{2}+t_{0}^{2})^{2}})}
+ t(t_{+}-t)\right]}
 & 0<t<t_{+} \\ \\
q\sim -\frac{(t-t_{+})^{6}\left(1 +\frac{8}{3h_{0}^{2}}\frac{t(t-t_{+})(t-t_{-})}{(t_{1}^{2}+t_{0}^{2})^{2}}\right)^{5}
\left(\frac{-(t_{rip}-t)(t_{rip}-t +2t_{+})}{(t_{rip}-t_{+})^{2}(1+\beta^{2}
+(t-t_{+})^{2})}+1\right)^{5}}{(t_{rip}-t)^{3}(t_{rip}-t +2t_{+})^{3}}  &
t_{+}<t<t_{rip} \\ \\
\end{array}
 \label{Q60}
\end{eqnarray}

\begin{figure*}[thbp]
\begin{tabular}{rl}
\includegraphics[width=5.5cm]{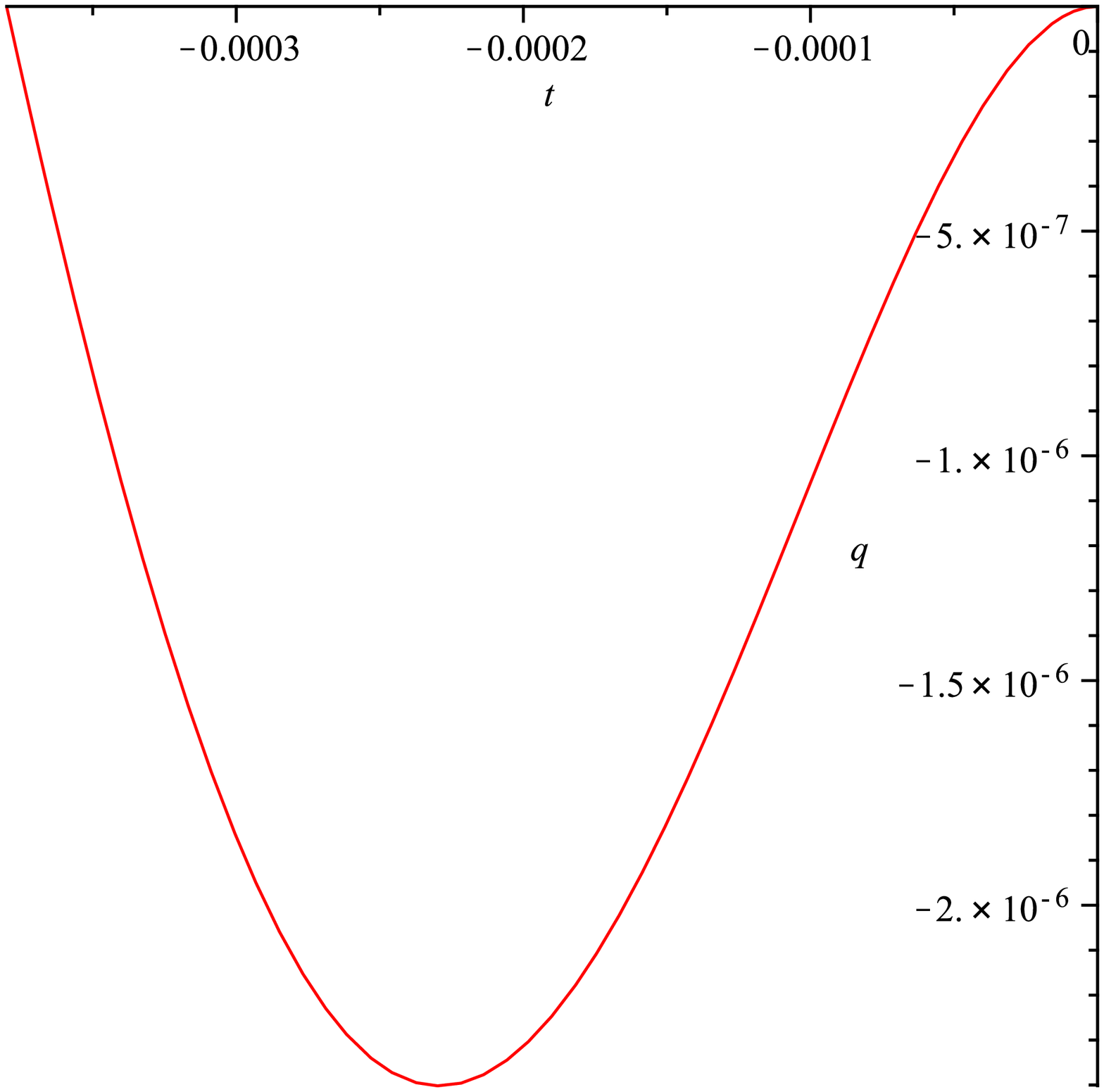}&
\includegraphics[width=5.5cm]{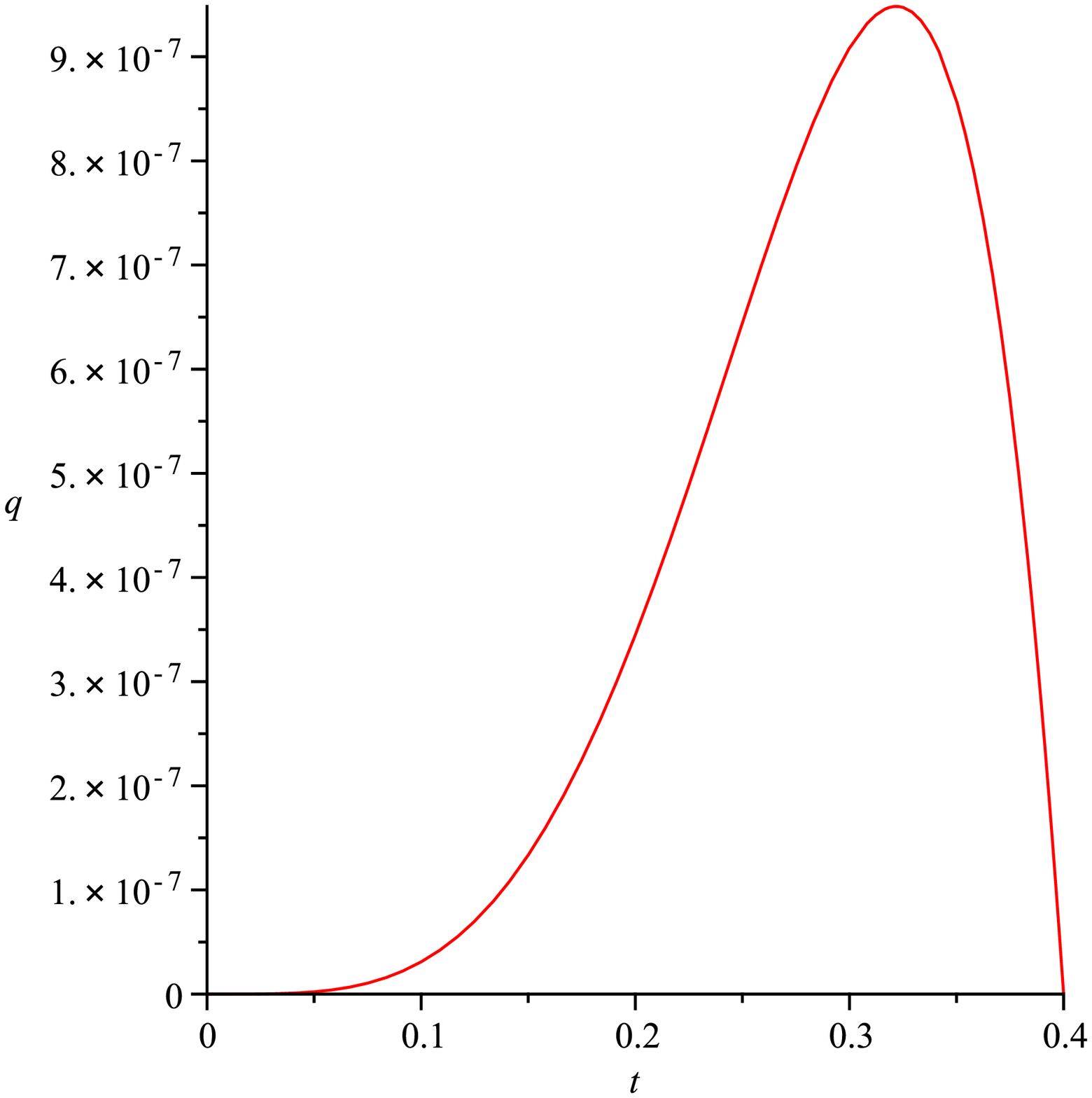}
\includegraphics[width=5.5cm]{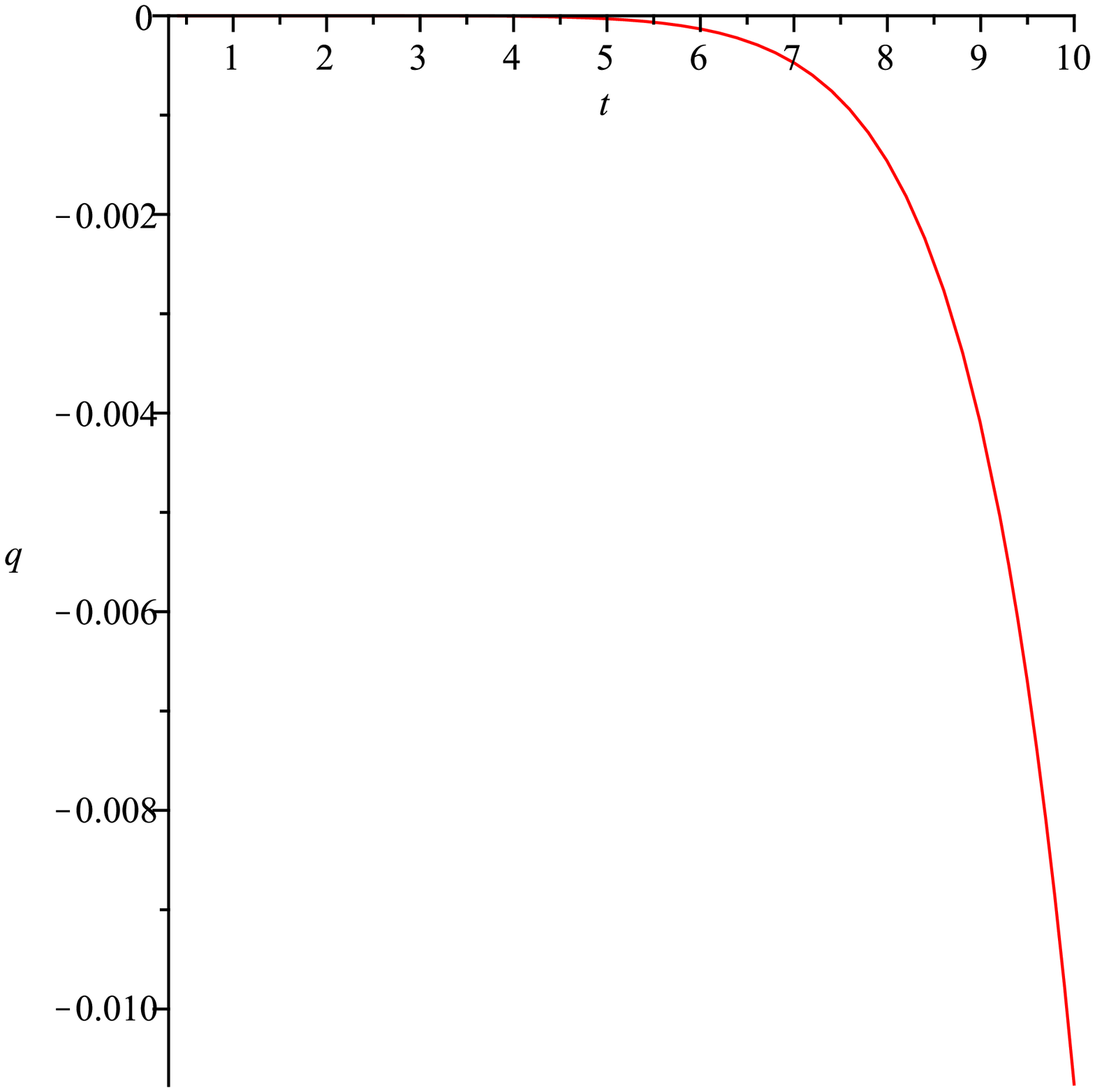}\\
\end{tabular}
\caption{ (1a Left) The deceleration  parameter  for inflation era
of expansion history as a function of the t where t is the age of
universe. (1b Middle)  The deceleration parameter for deceleration
era of expansion history as a function of the t where t is the
age of universe. (1c Right) The deceleration parameter  for late
time acceleration era of expansion history as a function of the t
where t is the age of universe. }
\end{figure*}


In Figures 1a,1b and 1c, we sketch the deceleration  parameter
for three phases of expansion history as a function of  the age of universe $t$. In these plots, we choose
$t_{-}=-0.005(yr)$, $t_{+}=0.4(Gyr)$ and $t_{rip}=30(Gyr)$. We
find that $q=-0.542$ leads to $t_{universe}= 13.5(Gyr)$. This result
is compatible with SNeIa data \cite{q22}. As it can be seen from
Fig. 1a, the deceleration parameter is negative in the range
$t_{-}<t<0$ and becomes zero at $t=0$. This means that the Universe
inflates in this period of time. In Fig. 1b, we observe that $q$
is zero at t=0 and $t=t_{+}$ and has a maximum in this epoch.
Finally, this parameter (Fig. 1c) is negative again in today
acceleration epoch and tends to $-\infty$ at Big Rip singularity.

\section{Summary and Discussion} \label{sum}
In this paper, we proposed a  model that allows to account for
dynamics of the transition from the phantom inflationary to the
non-phantom standard cosmology and to recover the today
observed acceleration epoch. At the first stage of evolution,
a BIon system  is  formed due to the dynamics of  black fundamental strings at
transition point. This BIon is a configuration in flat space of a
universe brane  and a parallel anti-brane connected by a
wormhole. With decreasing temperature, wormhole  becomes thinner,
its energy flows into the universe branes and causes their growth.
After a short time, this wormhole evaporates, inflation ends and non-phantom
era begins. Eventually, two universe brane and antibrane become
close to each other, tachyonic potential between them increases
and a  new wormhole is formed. In this condition, the Universe
evolves  from non-phantom phase to phantom one and consequently,
a phantom-dominated era of the Universe  accelerates and ends up into
Big-Rip singularity. Comparing this model with previous unified
cosmology models and  observational data, it is possible to obtain some
phenomenological parameters in terms of time.
In a forthcoming paper, we will develop the model in view of cosmological observations adopting the  approach discussed in \cite{q9}.

\section*{Acknowledgments}
\noindent
A. Sepehri would like to thank of Shahid Bahonar University of Kerman for financial support during
investigation in this work. He also thanks Prof. Harmark for his guidance. FR and AP wish to thank
the authorities of the Inter-University Centre for Astronomy and Astrophysics, Pune, India for providing
the Visiting Associateship. The financial supported by the UGC, India under the grant Project F.No.
41-899/2012 (SR) is gratefully acknowledged by AP. IHS is also thankful to DST, Govt. of India for providing
financial support under INSPIRE  Fellowship. S. Capozziello is supported by INFN ({\it iniziativa specifica} QGSKY).

 \end{document}